%% file: paper.tex
\documentclass[conference]{IEEEtran}
\IEEEoverridecommandlockouts

\usepackage[utf8]{inputenc}
\usepackage[T1]{fontenc}
\usepackage[english]{babel}

\usepackage{xcolor}
\usepackage{amsmath, amsfonts}

\usepackage{multirow}
\usepackage{blkarray}
\usepackage{makecell}

\usepackage{graphicx}
\graphicspath{{./images/}}

\usepackage[printonlyused, withpage, nolist]{acronym}
\usepackage{booktabs}
\usepackage{relsize}
\usepackage{hyperref}
\hypersetup{hidelinks, colorlinks=false, pdftitle={Evaluating SAP Joule for Code Generation}, pdfauthor={Joshua Heisler, Johannes Reisinger}}

\usepackage{todonotes}
\setuptodonotes{inline}

\usepackage{csquotes}
\usepackage[backend=biber, style=ieee, url=false, maxbibnames=99]{biblatex}
\def\BibTeX{{\rm B\kern-.05em{\sc i\kern-.025em b}\kern-.08em
T\kern-.1667em\lower.7ex\hbox{E}\kern-.125emX}}

\begin{document}
\title{Evaluating SAP Joule for Code Generation}

\author{
    \IEEEauthorblockN{Joshua Heisler}
    \IEEEauthorblockA{\textit{Faculty of Computer Science} \\
        \textit{Deggendorf Institute of Technology}\\
        Deggendorf, Germany \\
        paper@joshuaheisler.de}
    \and
    \IEEEauthorblockN{Johannes Reisinger}
    \IEEEauthorblockA{\textit{Faculty of Computer Science} \\
        \textit{Deggendorf Institute of Technology}\\
        Deggendorf, Germany \\
        johannes.reisinger@th-deg.de}
    \and
    \IEEEauthorblockN{Andreas Fischer}
    \IEEEauthorblockA{\textit{Faculty of Computer Science} \\
        \textit{Deggendorf Institute of Technology}\\
        Deggendorf, Germany \\
        andreas.fischer@th-deg.de}
}

\maketitle

\begin{abstract}
    SAP has released its own proprietary generative model SAP Joule,
    intended for various generative tasks, including serving as a code
    assistant for software engineers.
    While Joule is yet not focused on SAP-specific ABAP code generation,
    it can be used for other common languages, including Javascript.
    This paper compares SAP Joules Javascript
    coding capabilities against a total of 29 other models using the HumanEval-X Javascript benchmark.
    SAP Joule achieves a strict accuracy of 80.49\% as the fifth best model in
    our evaluation.
    To the best of our knowledge, this is the first comparative evaluation
    of SAP Joule code generation capabilities.
\end{abstract}


\input{sections/introduction}
\input{sections/related-work}

\input{sections/method}
\input{sections/results}
\input{sections/conclusion}

\medskip
\renewcommand*{\bibfont}{\footnotesize}
\setlength{\biblabelsep}{\labelsep}
\setlength{\bibitemsep}{\IEEEbibitemsep}
\printbibliography

\input{sections/acronyms}

\end{document}

%% file: sections/introduction.tex
\section{Introduction}\label{sec:introduction}

The advent of powerful transformer-based~\acp{llm} in recent years has marked a paradigm shift, with numerous publications demonstrating substantial success in code generation.
This progress is largely attributed to the vast quantities of publicly accessible code and associated documentation.

Nonetheless, a considerable segment of the software industry relies on proprietary codebases and specialized frameworks,
such as those within the SAP ecosystem.
These environments often utilize languages like ABAP and employ frameworks such as SAPUI5 or the~\ac{cap} model and
open source data is sparse.
Recognizing the burgeoning interest in \ac{llm} based code generation, SAP has introduced its own model SAP Joule.
It is intended for integration across various SAP products and services.
An early version of SAP Joule has been made available as a code generation assistant within SAP's cloud-based
development environment, Business Application Studio (BAS), which only supports the generation of JavaScript
code for SAP's Node.js-based CAP framework, which is utilized for creating REST-based backends for the SAPUI5 frontend framework.
At the time of this evaluation, SAP Joule does not officially support ABAP code generation, but this is expected to follow.

This paper aims to compare the JavaScript code generation capabilities of this initial release of SAP Joule with other prominent LLMs.
A significant constraint encountered during this investigation is that, at the time of research,
SAP Joule was accessible only through its integrated UI within BAS and lacked an API.
This limitation precludes the use of standard automated evaluation metrics for~\ac{llm}-based code generation,
such as pass@k, due to the prohibitive amount of manual interaction required to generate an unbiased and statistically significant dataset.
Similarly, the application of standard benchmark datasets is rendered impractical.

This contraint dictates the use of a metric that yields feasible and indicative results from a minimal number of manual queries.
Strict accuracy, a metric that measures the percentage of programming problems for which the generated solution successfully passes all test cases,
was identified as the most suitable candidate. While this metric has historically been viewed with caution due to a high variance in results
and therefore a possible bias, we argue that its application is justified for this specific investigation.
\citeauthor{hendrycksMeasuringCodingChallenge2021} suggested its potential viability for validation purposes
as~\acp{llm} achieve greater proficiency in code generation~\cite{hendrycksMeasuringCodingChallenge2021}. 


The main contribution of this paper is the evaluation of SAP Joule's JavaScript code generation capability using the strict accuracy
metric on the HumanEval-XL JavaScript benchmark, and comparing it to 29 recent models.

Following this introduction, Section~\ref{sec:related-work} will present related work.
Section~\ref{sec:methodology} will detail the methodology employed for the manual evaluation of SAP Joule and the
application of the strict accuracy metric.
The results of this evaluation will be presented and discussed in Section~\ref{sec:results}.
Section~\ref{subsec:compared-large-language-models} will compare these results with the performance of other
code generation models on various programming languages.
Finally, Section~\ref{sec:conclusion} will conclude our findings and suggest avenues for future research.

%% file: sections/related-work.tex
\section{Related Work}\label{sec:related-work}

\citeauthor{jiangSurveyLargeLanguage2024} present a comprehensive survey of large language models for code generation, offering a thorough examination
of various \acp{llm}~\cite{jiangSurveyLargeLanguage2024}.
Their survey facilitates a comparative analysis using the benchmark HumanEval,
evaluating 164 handwritten programming tasks for code generation in Python with pass@1.
The top performing closed source model is Claude-3.5-Sonnet with 92.0\% pass@1,
followed by the best open-source performing model DeepSeek-Coder-V2-Instruct with 90.2\%.
The average pass@1 performance for the top 5 performing models is 88.78\% for closed source models
and 84.62\% for open source models.

\citeauthor{houLargeLanguageModels2024a} conduct a systematic literature review on 395 publications regarding~\acp{llm} in~\ac{se}
classified 731 selected papers published in 2023 and 2024.
113 papers (15.5\%) regard an encoder-only architecture, 109 papers (14.9\%) regard an encoder-decoder architecture,
and 509 papers (69.6\%) regard a decoder-only architecture~\cite{houLargeLanguageModels2024a}.
The three major tasks in selected publications are software development (56.55\%), software maintenance (22.71\%),
and software quality assurance (15.14\%).

The HumanEval-X benchmark was published in union with the the CodeGeeX model~\cite{zhengCodeGeeXPreTrainedModel2023}.
The original HumanEval benchmark prompts, canonical solutions, and test cases were translated from Python to
C++, Java, JavaScript and Go, which the CodeGeeX model was trained on.

To the best of our knowledge, no other publication has evaluated SAP Joule~\cite{IntroducingJoule} for code generation capabilities
at the time of this study.

%% file: sections/method.tex
\section{Methodology}\label{sec:methodology}

To systematically assess the code generation capabilities of SAP Joule relative to other~\acp{llm},
this paper defines and executes a quantitative evaluation focused on the generation of JavaScript code.
This section outlines the framework for this comparison, detailing the selection of benchmark datasets,
evaluation metrics, the models for comparison, and the process used for evaluation.

\subsection{Datasets}\label{subsec:dataset}
Considering the need of manual prompting for SAP Joule, benchmarks with a manageable number of problems
in javascript are prioritized.

Due to a lack of pure JavaScript datasets, we review a number of multilingual benchmarks,
including HumanEvalPack~\cite{muennighoffOctoPackInstructionTuning2023},
HumanEval-X~\cite{zhengCodeGeeXPreTrainedModel2023},
Multilingual HumanEval and MBPX~\cite{athiwaratkunMultilingualEvaluationCode2022},
MultiPL-E~\cite{cassanoMultiPLEScalablePolyglot2023},
and xCodeEval~\cite{khanXCodeEvalLargeScale2023}.

HumanEval-X and Multilingual HumanEval, both containing 164 coding problems derived from the HumanEval benchmark, meet this criteria.
The HumanEval-X benchmark is selected, with a particular focus on its subset HumanEval-JS.

\subsection{Metrics}\label{subsec:metrics}

A number of metrics have been discussed in the literature for evaluation of \ac{llm}-based code assistants.
They can be divided into reference-based and execution-based metrics.

\subsubsection{Reference-based Metrics}\label{subsubsec:reference-based-metrics}

This type of metric uses a predefined reference solution as gold standard. 
The generated code is compared against this reference solution.
The distance to the reference solution is then used as a measure for the quality
of the code generator. Metrics following this approach include 
exact match, BLEU~\cite{papineniBLEUMethodAutomatic2001}, CodeBLEU~\cite{renCodeBLEUMethodAutomatic2020},
ROGUE~\cite{linROUGEPackageAutomatic2004} and METEOR~\cite{banerjeeMETEORAutomaticMetric2005}.
It should be noted that preceding research
identified an anticorrelation between reference-based scores and functional correctness~\cite{hendrycksMeasuringCodingChallenge2021}.
This is (partly) due to the fact that multiple implementations can yield the same correct functional outcome,

\subsubsection{Execution-based Metrics}\label{subsubsec:execution-based-metrics}
The second type of metric measures code quality by executing the generated
code and testing it with an extensive suite of unit tests. 
This approach, which directly assesses functional correctness via strict accuracy,
is the de-facto standard in code evaluation in recent publications of new code generating \acp{llm}.
Metrics in this area include test-case average and strict accuracy~\cite{hendrycksMeasuringCodingChallenge2021},
pass@k~\cite{chenEvaluatingLargeLanguage2021}, and n@k~\cite{liCompetitionlevelCodeGeneration2022}.

\subsubsection{Choosing strict accuracy as metric}

The HumanEval benchmark provides the basis for this evaluation. At the time of writing (June 2025) SAP Joule lacks API access and manual prompts are necessary.
The evaluation of the HumanEval benchmark is to be conducted by validating functional correctness. Consequently, execution-based metrics are more appropriate for the approach presented here.
The necessary manual effort renders metrics that demand a large number of generated samples per problem impractical.

Strict Accuracy is defined as the percentage of programming problems for which the single generated code solution passes all associated unit tests successfully.
While a possibility of receptiveness to single-run variance, i.e.~bias, exists, the utilization of this approach is supported
by the substantial progress in the generative capabilities of modern~\acp{llm} compared to the initial introduction
of execution-based metrics~\cite{hendrycksMeasuringCodingChallenge2021}.
Consequently, strict accuracy is selected as the primary evaluation metric.

\subsection{Compared Large Language Models}\label{subsec:compared-large-language-models}
A selection of relevant~\acp{llm} available in late 2024 is collated for the purpose of comparison, including SAP Joule.
The selection criteria comprise the following:
\begin{itemize}
    \item The models are released in 2023 or later
    \item The models are publicly available
    \item Both proprietary and open-source models are included
    \item The feasibility of executing open-source models on the available hardware is assessed, with a memory of 25GB
    \item The latest available version of each model is utilized
\end{itemize}

The evaluated models comprise the following (cf.~Table~\ref{tab:humanevaljs_results}):

\subsubsection{Proprietary}\label{subsubsec:proprietary}
SAP Joule, the OpenAI models GPT-4 Turbo, GPT-4o and GPT-3.5 Turbo, as well as the Anthropic models Claude 3.5 Sonnet,
Claude 3 Opus and Claude 3 Haiku.

\subsubsection{Open-Source}\label{subsubsec:open-source}
Qwen2.5
and Quen2.5 Coder,
DeepSeek Coder V2,
Codestral,
Gemma2,
Yi-Coder,
Mistral Small and Nemo,
WizardCoder,
CodeGeeX4,
CodeLlama,
CodeGemma,
the IBM Granite models,
OpenCodeInterpreter and the Llama series,
StarCoder2, and
Phi-3.5.

\subsection{Evaluation Process}\label{subsec:evaluation-process}
This section gives a detailed breakdown of the process for model preparation via prompting, model interaction, and response cleanup.

\subsubsection{Prompting}\label{subsubsec:prompting}
Previous tests indicate that~
Multilingual \acp{llm} are prone to change the language when not prompted carefully.
Indeed, this transfers to code generation, as well.
Tests indicate that models often do not respond in JavaScript but use other well-known programming languages such as Python.
The task, as defined in the benchmark, may be ambiguous with respect to the programming language to be used.
Consequently, a model may have issues accurately determining the correct language.
As a countermeasure, the prefix “Use JavaScript.” is appended to each prompt from the dataset
prior to its transmission to the~\acp{llm} to ensure consistent language generation.

\subsubsection{Model Interaction}\label{subsubsec:model-interaction}
Proprietary models are queried using their respective Batch APIs for cost-efficiency and volume handling.
Open-source models are executed locally using Ollama~\cite{Ollama} and accessed via its local API.
SAP Joule is manually prompted via its user interface, which is integrated into the cloud-based IDE SAP Build Code,
as there is no API available.

\subsubsection{Response Processing}\label{subsubsec:response-processing}
\Ac{llm} responses frequently incorporate natural language explanations, code block formatting, and occasionally incomplete code snippets or superfluous example calls.
To maintain fidelity to the original prompts of the benchmark, no additional specifications regarding the output are provided.
Therefore, the following standardized cleaning process is applied:

\begin{itemize}
    \item The removal of all non-code text, that is to say, all explanations and justifications.
    \item The removal of all code block markdown formatting.
    \item The elimination of any instance of function calls or the generation of test cases.
    \item Obtaining only the first single, complete code solution, despite the presence of multiple solutions offered by the model.
    \item Ensuring the presence of complete functions, by adding function declarations or closing brackets if necessary.
\end{itemize}

All this is done using a python script followed by a manual verification of each code snippet.

\subsubsection{Execution and Evaluation}\label{subsubsec:execution-and-evaluation}
For each model and each problem, the cleaned JavaScript code is concatenated with the corresponding test code from the dataset.
The execution of this combined script is facilitated by Node.js (v18 or later implied).
The Strict Accuracy calculation is deemed correct only if the Node.js process terminates without errors.
The errors detected include failed assertions, syntax errors, runtime errors, and timeouts (indicative of infinite loops).
The final Strict Accuracy score for each model is determined as the percentage of the 164 problems that are marked as correct.
The script for this evaluation is publicly available \cite{joshuaheislerdeRepository}.

%% file: sections/results.tex
\section{Results}\label{sec:results}

The quantitative evaluation yields strict accuracy scores for 30 \acp{llm} on the HumanEval-JS benchmark.

\begin{table}[ht]
    \caption{Results of the Evaluation of HumanEval-JS~\cite{zhengCodeGeeXPreTrainedModel2023}}
    \label{tab:humanevaljs_results}
    \centering
    \begin{tabular}{@{}llcr@{}}
        \toprule
        \textbf{Model}               & \textbf{Release} & \textbf{Strict Accuracy(\%)} & Ref                                                      \\
        \midrule
        claude-3-5-sonnet-20241022   & 2024-10          & 85.98                        & \cite{IntroducingClaude35}                               \\
        gpt-4o-2024-08-06            & 2024-05          & 85.98                        & \cite{IntroducingStructuredOutputs}                      \\
        gpt-4-turbo-2024-04-09       & 2023-11          & 84.15                        & \cite{ModelOpenAIAPI}                                    \\
        qwen2.5-32B                  & 2024-09          & 81.71                        & \cite{QwenQwen2532BHugging2025}                          \\
        \textbf{Joule (2024-10)}     & \textbf{2024-01} & \textbf{80.49}               & \cite{IntroducingJoule}                                  \\
        claude-3-opus-20240229       & 2024-02          & 78.05                        & \cite{ModelsOverview}                                    \\
        DeepSeek-Coder-V2-16b        & 2024-06          & 75.61                        & \cite{DeepseekaiDeepSeekCoderV22025}                     \\
        Qwen2.5-Coder-7B-Instruct    & 2024-09          & 73.78                        & \cite{QwenQwen25Coder7BInstructHugging2025}              \\
        Codestral-22B-v0.1           & 2024-05          & 73.78                        & \cite{MistralaiCodestral22Bv01Hugging}                   \\
        gemma2-27b                   & 2024-06          & 72.56                        & \cite{GoogleGemma227bHugging2025}                        \\
        gpt-3.5-turbo-0125           & 2023-03          & 70.73                        & \cite{OpenAIsGPT35Turbo}                                 \\
        Yi-Coder-9B                  & 2024-08          & 67.68                        & \cite{01aiYiCoder2025}                                   \\
        claude-3-haiku-20240307      & 2024-03          & 67.07                        & \cite{ModelsOverviewa}                                   \\
        mistral-small-22b            & 2024-09          & 65.24                        & \cite{BartowskiMistralSmall22BArliAIRPMaxv11GGUFHugging} \\
        WizardCoder-33B              & 2023-06          & 64.63                        & \cite{WizardLMTeamWizardCoder33BV11Hugging}              \\
        codegeex4-9b                 & 2024-07          & 62.80                        & \cite{THUDMCodegeex4all9bHugging2025}                    \\
        Phind-CodeLlama-34B-v2       & 2023-08          & 60.37                        & \cite{PhindPhindCodeLlama34Bv2Hugging}                   \\
        codegemma-7b                 & 2024-04          & 56.10                        & \cite{GoogleCodegemma7bHugging2025}                      \\
        mistral-nemo-12b             & 2024-07          & 54.27                        & \cite{NvidiaMistralNeMo12BBaseHugging}                   \\
        granite-3.0-8b-instruct      & 2024-10          & 51.22                        & \cite{IbmgraniteGranite308bInstruct2025}                 \\
        OpenCodeInterpreter-DS-33B   & 2024-02          & 50.61                        & \cite{MapOpenCodeInterpreterDS33BHugging2024}            \\
        Llama-3-8B-Instruct          & 2024-04          & 47.56                        & \cite{MetallamaMetaLlama38BInstructHugging2024}          \\
        starcoder2-15b-instruct-v0.1 & 2023-04          & 46.95                        & \cite{BigcodeStarcoder215binstructv01Hugging}            \\
        Llama-3-8B                   & 2024-04          & 46.34                        & \cite{MetallamaMetaLlama38BHugging2024}                  \\
        Llama-3.2-3B                 & 2024-09          & 42.07                        & \cite{MetallamaLlama323BHugging2024}                     \\
        llama3.2-3b-instruct-q8\_0   & 2024-09          & 40.24                        & \cite{Llama323binstructq8_0}                             \\
        Phi-3.5-3.8b                 & 2024-08          & 31.10                        & \cite{MicrosoftPhi35miniinstructHugging2025}             \\
        CodeLlama-34b-hf             & 2023-08          & 29.88                        & \cite{CodellamaCodeLlama34bHf}                           \\
        CodeLlama-7b-hf              & 2023-08          & 28.66                        & \cite{CodellamaCodeLlama7bHf}                            \\
        granite-34b-code-instruct-8k & 2024-04          & 18.90                        & \cite{IbmgraniteGranite34bCodeinstruct8k2023}            \\
        \bottomrule
    \end{tabular}
\end{table}

\begin{figure}[]
    \centering
    \includegraphics[width=1.0\linewidth]{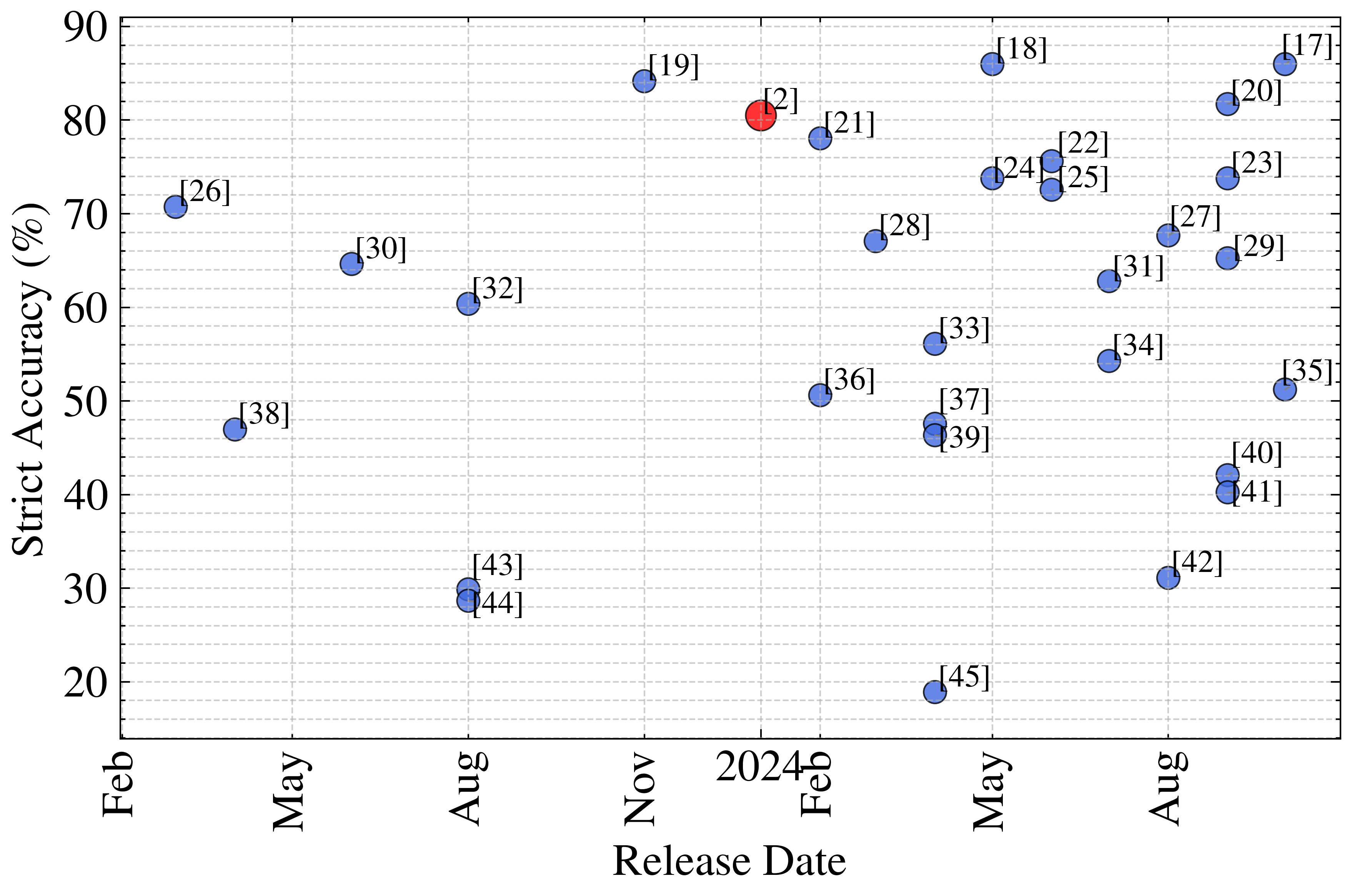}

    \caption{Measured strict accuracy performance over publication date of the investigated models. While
    even recent models vary wildly regarding strict accuracy performance, a slight trend towards better models can be observed.}
    \label{fig:llm-preformance-over-parameters}
\end{figure}

\subsection{Overall Performance Landscape}\label{subsec:overall-performance-landscape}
Across the range of evaluated models, the average strict accuracy is 60.6\%.
However, a considerable degree of variation is observed, as indicated by a standard deviation of 19\%.
This finding suggests a broad spectrum of capabilities among modern LLMs in terms of code generation with the JavaScript language (cf.~Fig.~\ref{fig:llm-preformance-over-parameters}).
The average performance figures demonstrate advancement in the domain of automated code generation.

\subsection{Performance of Leading Models}\label{subsec:performance-of-leading-models}
Leading proprietary models achieve the highest levels of accuracy: Anthropic’s Claude 3.5 Sonnet and OpenAI’s GPT-4o
achieve a tie for first position with a strict accuracy of 85.98\%,.
OpenAI’s GPT-4 Turbo demonstrates a closely related result of 84.15\%.
SAP Joule achieves an 80.5\% strict accuracy score.
This outcome positions it within the top 5 models in its field.
The open-source model Qwen2.5 achieves an 81.7\% accuracy,
a figure that surpasses the performance of both SAP Joule and Claude 3 Opus,
establishing it as the leading open-source model in this evaluation.

\subsection{Comparing SAP Joule with Open-Source Model Capabilities}\label{subsec:open-source-model-capabilities}

Several open-source contenders exhibit notable results.
Beyond Qwen 2.5, models such as DeepSeek-Coder-V2, Qwen2.5-Coder, Codestral, and Gemma2 all achieve strict accuracy scores above 70\%.
The performance of models specifically trained or fine-tuned for code is evident, 
sometimes even with fewer parameters (e.g., Qwen2.5-Coder with 7B parameters attaining approximately 74\% strict accuracy).
This suggests that specialized models can effectively compete with larger, generic models.

\begin{figure}[]
    \centering
    \includegraphics[width=1.0\linewidth]{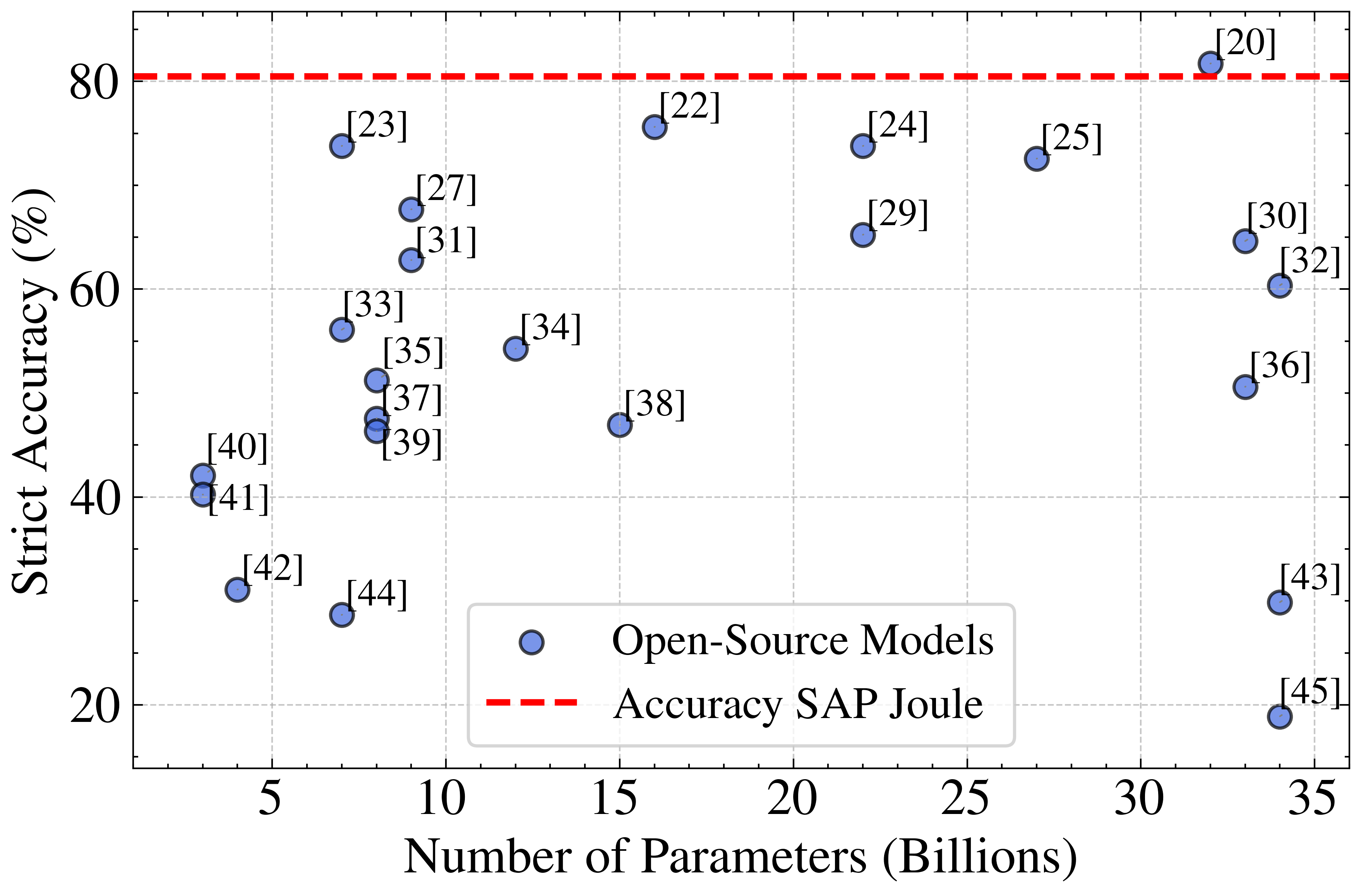}
    \caption{Performance in relation to the number of parameters for open source models. Since SAP Joule
    parameter count is unknown, the strict accuracy result is depicted independent of the parameter count.}
    \label{fig:sa-parameters}
\end{figure}

An analysis plotting strict accuracy against the number of parameters for the open-source models reveals only a weak correlation
($r = 0.23$), suggesting the relationship is possibly insignificant (cf.~Fig.~\ref{fig:sa-parameters}).

The performance appears to be more influenced by factors such as the model architecture, the training data, and specialization than by the parameter count alone within the evaluated range.

SAP Joule, being closed-source, does not disclose information regarding its model parameters.
Nevertheless, our evaluation shows that it is capable of producing reasonable results in comparison
to those of high-performing models---both proprietary and open-source.

%% file: sections/conclusion.tex
\section{Conclusion}\label{sec:conclusion}

We evaluated the SAP Joule model for JavaScript code generation capabilities
and compared it to 29 recent code generation models with less than 25GB memory footprint
on the HumanEval-X JavaScript dataset.
We used strict accuracy as metric for its robust evaluation of code functionality.
In our evaluation, SAP Joule achieved a result of 80.49\%---the fifth best result in the test.
Apart from proprietary models, only the best open source model gwen2.5-32b outperformed
SAP Joule. All contenders are within 6 percentage points (80\%-86\%).
To the best of our knowledge, this is the first
quantitative evaluation of SAP Joule's coding assistant capabilities.

Considering that SAP Joule is marketed as a general-purpose Large Language Model, 
which is primarily intended to support business processes in various SAP applications and 
is only available in a few developer-centric tools, its performance in the domain of code
generation is remarkable.

As SAP Joule has also been available for ABAP development since the beginning of Q2 2025, 
a future evaluation of its code generation capabilities with the ABAP programming language is warranted. 
In June 2025, SAP announced that similar internal evaluations concerning ABAP
had commenced~\cite{ABAPConf2025}. Thus, verifying the ABAP capabilities of SAP Joule is an
obvious next step.

%% file: sections/acronyms.tex
\begin{acronym}[AAAAAAA]
    \acro{llm}[LLM]{large language model}
    \acro{se}[SE]{software engineering}
    \acro{cap}[CAP]{cloud application programming}
\end{acronym}